\documentclass[aps,prl,showpacs,twocolumn,preprintnumbers,superscriptaddress,nofootinbib,floatfix,10pt]{revtex4-1}
\usepackage{amsfonts,amssymb,stmaryrd,latexsym,amsmath,braket}
\usepackage{graphicx,subfigure}
\usepackage{comment}
\usepackage{times}
\usepackage{slashed}
\usepackage{bm}
\usepackage{braket}

\newcommand{\beginsupplement}{%
        \setcounter{table}{0}
        \renewcommand{\thetable}{S\arabic{table}}%
        \setcounter{figure}{0}
        \renewcommand{\thefigure}{S\arabic{figure}}%
        \setcounter{equation}{0}
        \renewcommand{\theequation}{S\arabic{equation}}%
     }

\makeatletter
\let\saved@includegraphics\includegraphics
\AtBeginDocument{\let\includegraphics\saved@includegraphics}

\makeatother

\begin{document}

\title{Rodeo Algorithm for Quantum Computing}
\author{Kenneth Choi}

\affiliation{Ridgefield High School, Ridgefield, CT 06877, USA}

\author{Dean Lee}

\affiliation{Facility for Rare Isotope Beams and Department of Physics and
Astronomy,
Michigan State University, MI 48824, USA}

\author{Joey Bonitati}

\affiliation{Facility for Rare Isotope Beams and Department of Physics and
Astronomy,
Michigan State University, MI 48824, USA}

\author{Zhengrong Qian}

\affiliation{Facility for Rare Isotope Beams and Department of Physics and
Astronomy,
Michigan State University, MI 48824, USA}

\author{Jacob Watkins}

\affiliation{Facility for Rare Isotope Beams and Department of Physics and
Astronomy,
Michigan State University, MI 48824, USA}

\begin{abstract} 
We present a stochastic quantum computing algorithm that can prepare any eigenvector of a quantum Hamiltonian within a selected energy interval $[E-\epsilon, E+\epsilon]$.  In order to reduce the spectral weight of all other eigenvectors by a suppression factor $\delta$, the required computational effort scales as $O[|\log \delta|/(p \epsilon)]$, where $p$ is the squared overlap of the initial state with the target eigenvector. The method, which we call the rodeo algorithm, uses auxiliary qubits to control the time evolution of the Hamiltonian minus some tunable parameter $E$. With each auxiliary qubit measurement, the amplitudes of the eigenvectors are multiplied by a stochastic factor that depends on the proximity of their energy to $E$. In this manner, we converge to the target eigenvector with exponential accuracy in the number of measurements.  In addition to preparing eigenvectors, the method can also compute the full spectrum of the Hamiltonian.  We illustrate the performance with several examples.  For energy eigenvalue determination with error $\epsilon$, the computational scaling is $O[(\log \epsilon)^2/(p \epsilon)]$.  For eigenstate preparation, the computational scaling is $O(\log \Delta/p)$, where $\Delta$ is the magnitude of the orthogonal component of the residual vector.  The speed for eigenstate preparation is exponentially faster than that for phase estimation or adiabatic evolution.
\end{abstract}
\maketitle

Quantum computing is a powerful paradigm with the potential to describe large complex systems and eventually perform computations beyond the reach of classical computing. Recently, there have been several exciting algorithmic advances in describing the time evolution of Hamiltonians on quantum computers using a variety of different tools \cite{Low:2016a,Low:2016b,Childs:2018a,Campbell:2019a,Childs:2019b,Meister:2020a}.  They can be broadly categorized as either Lie-Trotter-Suzuki product formulas \cite{Trotter:1959,Suzuki:1976a} or linear combinations of unitaries \cite{Childs:2012a}.  Unfortunately, the application of these techniques for quantum state preparation is limited by existing hardware capabilities. Quantum adiabatic evolution is one approach to quantum state preparation that starts with an eigenstate of a simple Hamiltonian that slowly evolves with an interpolating time-dependent Hamiltonian until reaching the desired target Hamiltonian \cite{Farhi:2000a,Farhi:2001a}.  The problem is that calculations based on quantum adiabatic evolution require an extended time evolution that makes the computational cost prohibitive for large systems. To address this problem, we introduce a new framework for quantum state preparation and spectrum determination called the rodeo algorithm. 

The rodeo algorithm employs a strategy that is opposite to quantum adiabatic evolution.  As the name suggests, the rodeo algorithm operates by shaking off all other states until only the target eigenvector remains.  In this regard, the rodeo algorithm is similar in character to the projected cooling algorithm \cite{Lee:2019zze,Gustafson:2020vqg}.  However, the rodeo algorithm has the advantage that it can be applied to any quantum Hamiltonian and is a recursive algorithm that achieves exponential convergence in the number of cycles.  It can be used to compute the full energy spectrum as well as prepare any energy eigenstate. While the rodeo algorithm might appear similar to Kitaev's iterative version of quantum phase estimation \cite{Kitaev:1995} and fixed-time energy band filtering methods \cite{Ge:2017,Lu:2020}, none of these methods can be used efficiently to prepare individual eigenstates of a general quantum Hamiltonian.

We will refer to the Hamiltonian of interest as the object Hamiltonian, $H_{\rm obj}$, and the linear space which it acts upon the object system. By assumption, the object system starts in some initial state $\ket{\psi_I}$, which in general will update after each measurement. We will use auxiliary or ancilla qubits coupled to the object system.  In the following we use the standard terminology, ancilla qubits. But we also mention that this collection of ancilla qubits is also informally called the ``rodeo arena''.

\begin{figure}
\centering
\includegraphics[width=8.5cm]{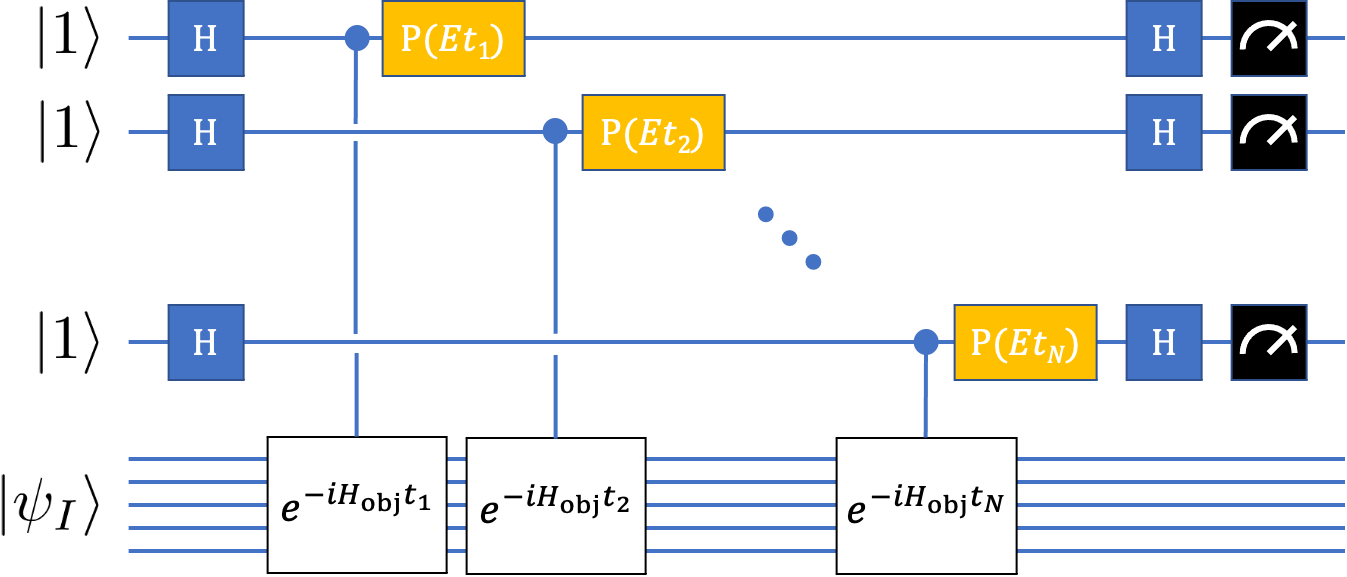}
\caption{(color online) {\bf Circuit diagram for the rodeo algorithm.} The object system starts in an arbitrary state $\ket{\psi_I}$.  Each of the ancilla qubits are initialized in the state $\ket{1}$ and operated on by a Hadamard gate H.  We use each ancilla qubit $n=1, \cdots, N$ for the controlled time evolution of the object Hamiltonian, $H_{\rm obj}$, for time $t_n$.  This is followed by a phase rotation P$(Et_n)$ on ancilla qubit $n$, another Hadamard gate H, and then measurement.}
\label{rodeo_circuit}
\end{figure}

If the quantum device allows for mid-circuit measurements, then only one ancilla qubit is needed.  However, here we focus on the implementation using different ancilla qubits for each cycle of the rodeo algorithm.  Each of the ancilla qubits is initialized in the state $\ket{1}$ and operated on by a Hadamard gate H.  We use each ancilla qubit $n=1, \cdots, N$ to control the time evolution of $H_{\rm obj}$ for time $t_n$.  In order to achieve the desired energy filtering, we operate on each ancilla qubit $n$ with the phase rotation gate P$(Et_n)$, follow that with another Hadamard gate H, and then measure the qubit.  We use the convention that P$(Et_n)$ multiplies the phase $e^{iEt_n}$ to the $\ket{1}$ state and leaves the $\ket{0}$ state untouched. The circuit diagram for the rodeo algorithm is shown in Fig.~\ref{rodeo_circuit}.

In order to illustrate the effect of these gate operations, let us explicitly write out the operation for one cycle of the rodeo algorithm with one ancilla qubit.  Starting from the initial state $\ket{1} \otimes \ket{\psi_I}$ and performing one rodeo cycle, we obtain
\begin{align}
& \begin{bmatrix}
\left[ \tfrac{I}{2}-\tfrac{1}{2}e^{-i(H_{\rm obj}-E)t_1} \right] \ket{\psi_I} \\
\left[ \tfrac{I}{2}+\tfrac{1}{2}e^{-i(H_{\rm obj}-E)t_1} \right] \ket{\psi_I}
\end{bmatrix} =  \nonumber \\
& \begin{bmatrix}
\tfrac{I}{\sqrt{2}} &
\tfrac{I}{\sqrt{2}}\\
\tfrac{I}{\sqrt{2}}& 
\tfrac{-I}{\sqrt{2}}
\end{bmatrix}
\begin{bmatrix}
 I & 0 \\
0 & Ie^{iEt_1} 
\end{bmatrix}
 \begin{bmatrix}
 I & 0 \\
0 & e^{-iH_{\rm obj}t_1}
\end{bmatrix} 
\begin{bmatrix}
\tfrac{I}{\sqrt{2}} & 
\tfrac{I}{\sqrt{2}}\\
\tfrac{I}{\sqrt{2}} & 
\tfrac{-I}{\sqrt{2}}
\end{bmatrix}
\begin{bmatrix}
0 \\
\ket{\psi_I}
\end{bmatrix},
\label{one_rodeo}
\end{align}
where $I$ is the identity operator on the object system.
We note that $H_{\rm obj}$ commutes with all of our gates, and so we can describe the action of the rodeo algorithm for each individual eigenvector of $H_{\rm obj}$ with energy $E_{\rm obj}$.  In that case, the probability of measuring the ancilla qubit $n$ in the $\ket{1}$ state is 
\begin{equation}
   \cos^2\left[(E_{\rm obj}
   -E)\tfrac{t_n}{2}\right] = \left\vert \tfrac{1}{2}+
   \tfrac{1}{2}e^{-i(E_{\rm obj}-E)t_n} \right\vert^2.
\end{equation}
The success probability of measuring all $N$ ancilla qubits in the $\ket{1}$ state is given by product, 
\begin{equation}
   P_N =\prod_{n=1}^N \cos^2\left[(E_{\rm obj}
   -E)\tfrac{t_n}{2}\right]. \label{prob}
\end{equation}
If we now take random values of $t_n$, we have an energy filter for $E_{\rm obj}=E$.  The geometric mean of $\cos^2\theta$ when sampled uniformly over all $\theta$ is equal to $\tfrac{1}{4}$.  Therefore the spectral weight for any $E_{\rm obj}\ne E$ is suppressed by a factor of $\tfrac{1}{4^N}$ for large $N$.

\begin{figure}
\centering
\includegraphics[width=8.5cm]{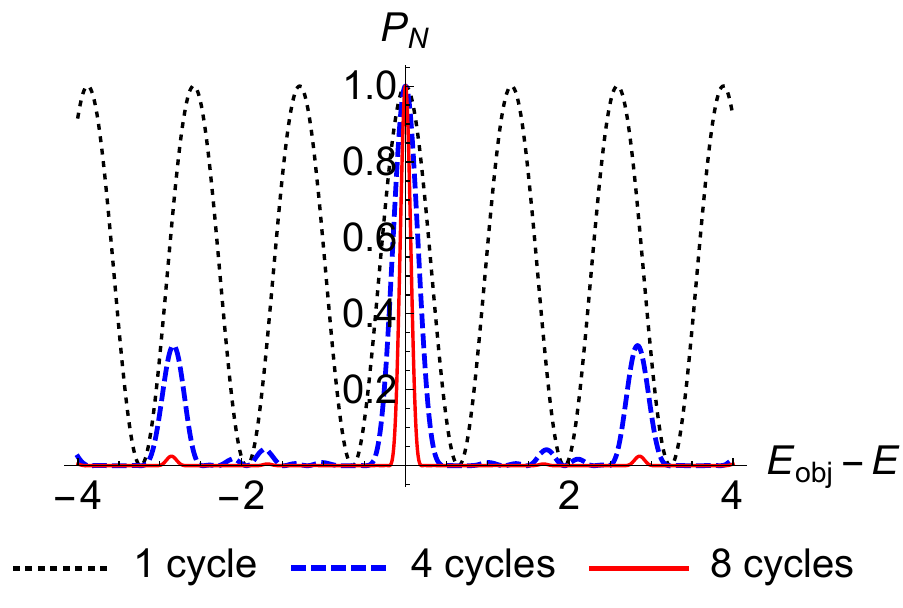}
\caption{(color online) {\bf Measurement probability.} We plot the probability $P_N$ of measuring the $\ket{1}$ state for all ancilla qubits versus $E_{\rm obj}-E$ for $1$ (dotted black line), $4$ (dashed blue line), and $8$ (solid red line) cycles with Gaussian random values of $t_n$ with $t_{\rm RMS}=10$.}
\label{rodeo_probability}
\end{figure}

In Fig.~\ref{rodeo_probability} we plot the probability $P_N$ of measuring the $\ket{1}$ state for all ancilla qubits versus $E_{\rm obj}-E$ for $1$ (dotted black line), $4$ (dashed blue line), and $8$ (solid red line) cycles with Gaussian random values of $t_n$ with root-mean-square value $t_{\rm RMS}=10$.  If we use a Gaussian approximation for $P_N$ near its maximum value of $1$ at $E_{\rm obj}=E$, we find that the width of the peak scales as $O[1/(\sqrt{N}t_{\rm RMS})]$.

Let $\epsilon$ be the desired energy resolution of our rodeo algorithm such that all energy eigenvectors outside of the interval $[E-\epsilon, E+\epsilon]$ are exponentially suppressed.  If we choose $t_{\rm RMS}$ to scale proportionally with $1/\epsilon$, then we achieve the desired energy filtering with energy resolution $\epsilon$.  The actual peak width will be a factor of $1/\sqrt{N}$ narrower than $\epsilon$, but that is needed to get exponential suppression as a function of $N$ for all energies $E_{\rm obj}$ further than $\epsilon$ from $E$.  

\begin{figure}
\centering
\includegraphics[width=8.5cm]{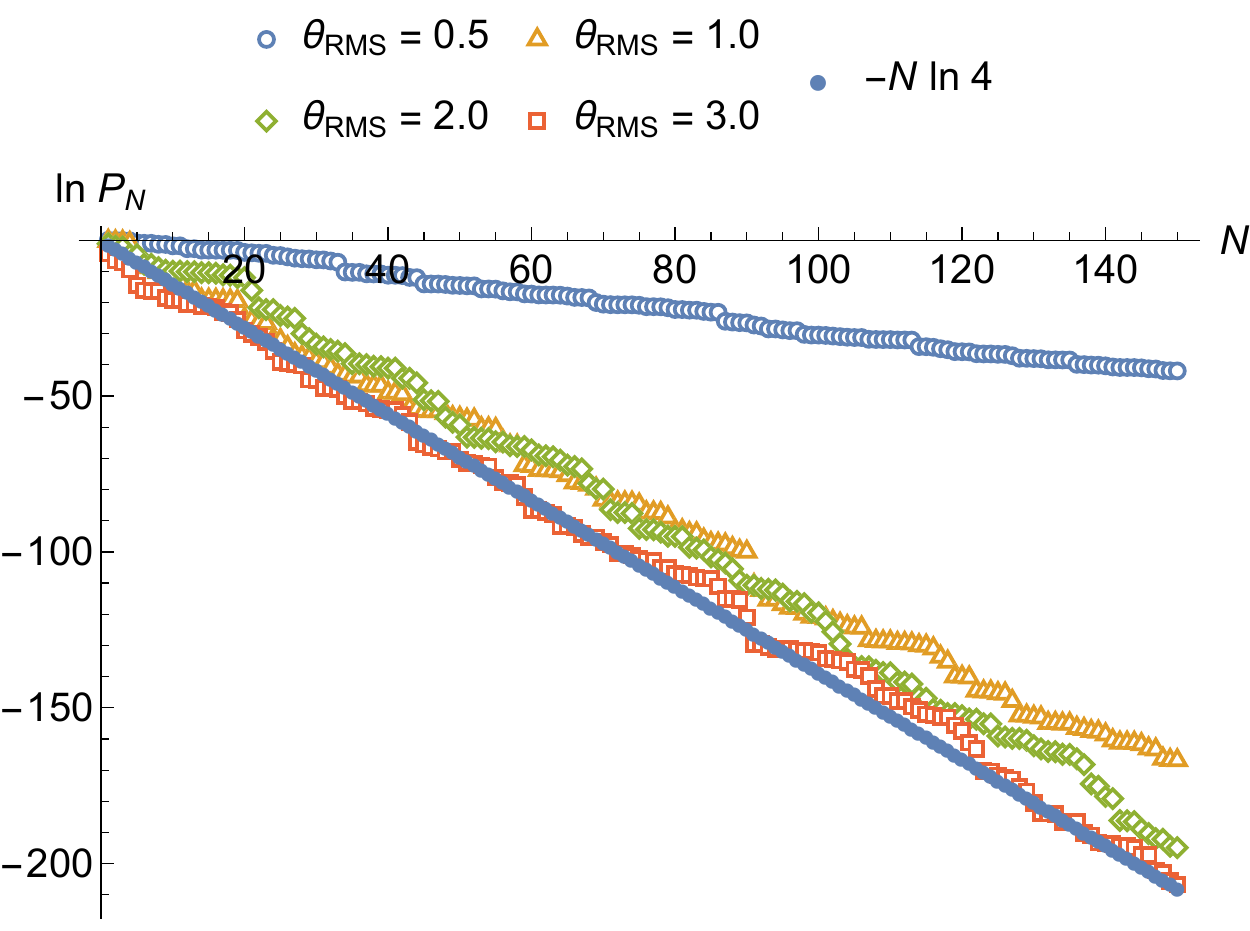}
\caption{(color online) {\bf Asymptotic scaling.} We plot $\ln P_N$ versus $N$ for Gaussian random values of $t_n$ with several selected values for $\theta_{\rm RMS}\equiv(E_{\rm obj}-E)\tfrac{t_{\rm RMS}}{2}$.  We show $\theta_{\rm RMS}=0.5$ (open circles), $\theta_{\rm RMS}=1.0$ (open triangles), $\theta_{\rm RMS}=2.0$ (open diamonds), and $\theta_{\rm RMS}=3.0$ (open squares). We show the predicted asymptotic scaling, $\ln P_N = -N \ln 4$, with filled circles.}
\label{asymptotic}
\end{figure}

In Fig.~\ref{asymptotic} we plot $\ln P_N$ versus $N$ for Gaussian random values of $t_n$ using several values for $\theta_{\rm RMS}\equiv(E_{\rm obj}-E)\tfrac{t_{\rm RMS}}{2}$.  We show $\theta_{\rm RMS}=0.5$ (open circles), $\theta_{\rm RMS}=1.0$ (open triangles), $\theta_{\rm RMS}=2.0$ (open diamonds), and $\theta_{\rm RMS}=3.0$ (open squares). We present the predicted asymptotic scaling, $\ln P_N = -N \ln 4$, with filled circles.  We see that for $\theta_{\rm RMS}$ greater than $1$, the expected asymptotic scaling is achieved.  Therefore, if $t_{\rm RMS}$ is larger than twice the inverse spacing between energy levels, then $P_N$ scales as $\tfrac{1}{4^N}$ for large $N$.

 Let us now consider what happens for an arbitrary initial state $\ket{\psi_I}$.  Let $E_j$ be the energy eigenvalue nearest to $E$, and let $\ket{E_j}$ be the corresponding eigenvector. In the limit of large $N$, the probability that we measure the $\ket{1}$ state $N$ times in a row is $pP_N$, where $p$ is the overlap probability of the initial state with $\ket{E_j}$, and $P_N$ is the success probability for $E_{\rm obj}=E_j$.  By tuning $E$ equal to $E_j$, this probability becomes $p$. If we require that the spectral weights of all other energy eigenvectors outside the interval $[E-\epsilon,  E+\epsilon]$ are suppressed by a factor $\delta$, then the computational effort scaling for the rodeo algorithm is $N t_{\rm RMS}/p = O[|\log \delta|/(p \epsilon)]$. 

In order to successfully determine any given energy eigenvalue $E_j$ with error $\epsilon$, the computational cost scales as $O[(\log \epsilon)^2/(p \epsilon)]$.  The search process involves $O(\log \epsilon)$ sequential scans of the energy, each scan sweeping over an energy range that is some constant factor $K$ smaller than the previous scan. Each scan is performed for several evenly spaced values of $E$, with a fixed number of rodeo cycles, and $t_{\rm RMS}$ a factor of $K$ larger than that used for the previous scan.  The total time evolution required will scale as $O(1/\epsilon)$,  and the factor of $(\log \epsilon)^2/p$ comes from the required statistics needed to perform the energy scans successfully with high probability. The resulting performance as a function of $\epsilon$ is close to the $O(1/\epsilon)$ bound set by the Heisenberg uncertainty principle.  For comparison, the computational effort for phase estimation is $O[1/(p \epsilon)]$ plus an additional cost that is $O[(\log \epsilon)^2]$ associated with the quantum Fourier transform.  Iterative phase estimation eliminates the need for the quantum Fourier transform, but is suitable only for finding the energy of a pure eigenstate.  We note that the direct calculation of the expectation value of the Hamiltonian for a pure eigenstate requires $O(1/\epsilon^2)$ measurements due to statistical errors.

In order to successfully prepare any given eigenstate $\ket{E_j}$ with a residual orthogonal component that has magnitude $\Delta$, the computational cost is $O(\log \Delta /p)$.  For this case, we keep $t_{\rm RMS}$ fixed but large enough that we are filtering out only the desired eigenstate.  We perform $N = O(\log \Delta)$ cycles of the rodeo algorithm, and this must be multiplied by $1/p$ for the number of measurements required.  In preparing the eigenstate, it is important to keep $E$ centered on the peak maximum associated with $E_j$.  Re-centering $E$ with each cycle requires only a constant overall factor in the computational cost that is independent of $p$ and $\Delta$.  In contrast, the computational cost for the same task using phase estimation requires an effort that scales as $O[1/(p \Delta)]$.  Adiabatic evolution requires an effort that is $O(1/\Delta)$ times a function of $p$ which depends on the adiabatic path connecting the initial and final Hamiltonians \cite{Wiebe:2011a}.  We see that for eigenstate preparation, the rodeo algorithm is exponentially faster than both phase estimation and adiabatic evolution in the limit $\Delta \rightarrow 0$.

 From Eq.~(\ref{prob}), we can also derive two useful estimates for $\Delta$, the magnitude of the residual orthogonal component, as a function of the number of rodeo cycles, $N$,
 \begin{align}
    F_A \equiv \sqrt{2^{-N}(1-p)/[p+2^{-N}(1-p)]},\nonumber \\
     F_G \equiv \sqrt{4^{-N}(1-p)/[p+4^{-N}(1-p)]}. \label{estimates}
 \end{align}
$F_A$ is appropriate for the case when $N$ is much smaller than the number of eigenstates with nonzero overlap with our initial state.  This corresponds with a spectral suppression factor of $1/2$ at each order for the undesired eigenstates, corresponding with the arithmetic mean of $\cos^2 \theta$.  $F_G$ is appropriate for the case when $N$ is much larger than the number of eigenstates with nonzero overlap with our initial state.  This corresponds with a spectral suppression factor of $1/4$ at each order for the undesired eigenstates, corresponding with the geometric mean of $\cos^2 \theta$.  

The rodeo algorithm depends heavily on the initial-state overlap probability $p$.  It is, therefore, very helpful to improve the quality of the initial state.  One possible approach is to make a variational ansatz based on physical intuition about the nature of the eigenvector.  Another strategy is to use domain decomposition to define a variational ansatz as a tensor product of wave functions on smaller subsystems.  Yet another approach is some combination of variational methods and adiabatic evolution such as the quantum approximate optimization algorithm \cite{Farhi:2014}.

As a first application of the rodeo algorithm, we consider the spin-$\tfrac{1}{2}$ Heisenberg model in a uniform magnetic field with $10$ sites forming a closed one-dimensional chain \cite{Heisenberg:1928}.  The Hamiltonian has the form
\begin{equation}
    H_{\rm obj} = J \sum_{\braket{j,k}} \vec{\sigma}_j\cdot \vec{\sigma}_k + h \sum_{j} \sigma^z_j,
\end{equation}
where $J$ is the exchange coupling, $\vec{\sigma}_j$ are the Pauli matrices on site $j$, $\braket{j,k}$ indicates nearest neighbors, and $h$ is the coupling to a uniform magnetic field in the $z$ direction.  We consider the antiferromagnetic case with values $J=1$ and $h=3$. For our initial state we use an alternating tensor product state,
\begin{equation}
    \ket{\psi_I} = \ket{0101010101}.
\end{equation}
Since our initial state has a high degree of symmetry, we expect our initial state to have nonzero overlap with a relatively small number of energy eigenstates.

\begin{figure}
\centering
\includegraphics[width=8.5cm]{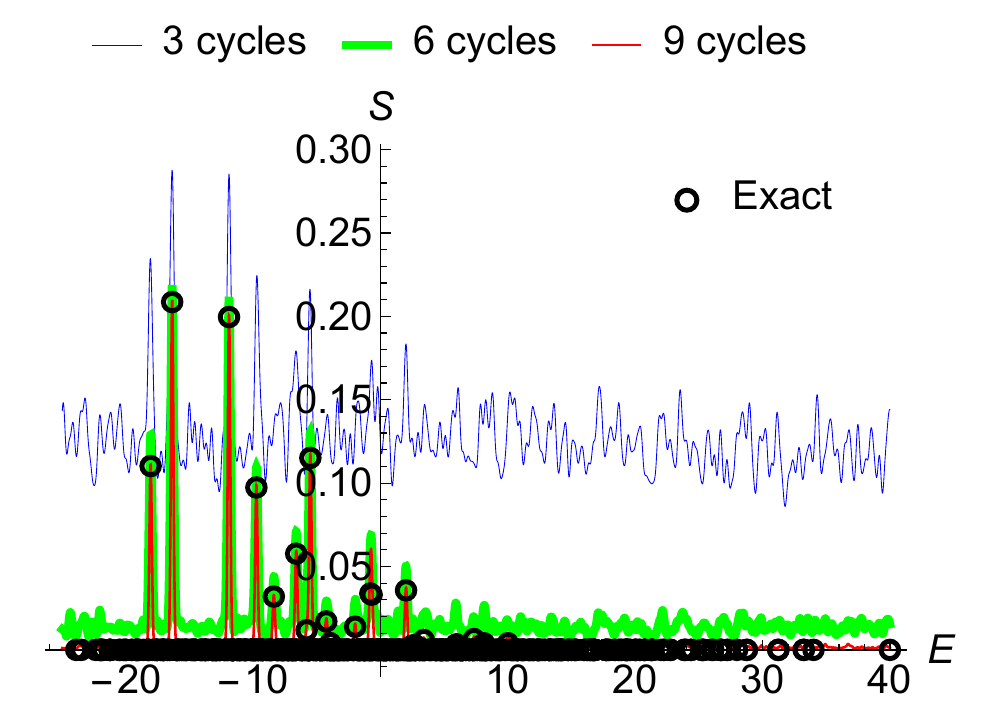}
\caption{(color online) {\bf Initial-state spectral function for the Heisenberg model.} We plot the initial-state spectral function using the rodeo algorithm for the Heisenberg spin chain with $3$ (thin blue line), $6$ (thick green line), and $9$ (medium red line) cycles. We have averaged over $20$ sets of Gaussian random values for $t_n$ with $t_{\rm RMS}=5$.  For comparison, we also show the exact initial-state spectral function with black open circles.}
\label{Heisenberg_spectrum}
\end{figure}

Let us label the energy eigenstates of $H_{\rm obj}$ as $\ket{E_j}$. We define the initial-state spectral function as $S(E)=|\braket{E_j|\psi_I}|^2$ for $E=E_j$ and $S(E)=0$ otherwise.  For the case of exact degeneracy, we sum the contribution from all degenerate energy states.  In Fig.~\ref{Heisenberg_spectrum}, we plot the initial-state spectral function using the rodeo algorithm for the Heisenberg spin chain with $N=3$ (thin blue line), $6$ (thick green line), and $9$ (medium red line) cycles. We have averaged over $20$ sets of Gaussian random values for $t_n$ with $t_{\rm RMS}=5$.  This averaging over sets of random values for $t_n$ decreases the stochastic noise and results in a roughly constant background that can be distinguished from the spectral signal.  For comparison, we show the exact initial-state spectral function with black open circles.  We see that the agreement obtained using the rodeo algorithm is excellent.  

The real challenge will be to perform these calculations on quantum computing devices with gate errors, measurement errors, and short decoherence times.  But it is promising that we can obtain good results even though neither $t_{\rm RMS}$ nor $N$ are very large.  The resulting short gate depth is crucial for implementation on noisy quantum devices.  In the Supplemental Materials, we show the corresponding results with $t_{\rm RMS}=1$.  Even in that case, we can clearly identify the spectrum of energy states with strong overlap with the initial state.

In addition to computing the initial-state spectral function, we can also prepare any energy eigenstate that has nonzero overlap with our initial state.  In the Supplemental Materials, we show the overlap probability with energy eigenvector $\ket{E_j}$ after $N$ cycles of the rodeo algorithm.  All of the energy eigenvectors with nonzero overlap with our initial state can be prepared with a relatively small number of rodeo cycles. After an energy eigenstate has been prepared using the rodeo algorithm, we can measure any properties of that eigenstate that we choose, such as expectation values of observables, transition matrix elements to other energy eigenstates, and linear response functions.  Several examples of such calculations using prepared energy eigenstates will be discussed in a future publication.  See also Ref.~\cite{Roggero:2018hrn,Roggero:2019myu,Roggero:2020qoz} for some approaches to computing linear response functions and spectral densities.

\begin{figure}
\centering
\includegraphics[width=8.5cm]{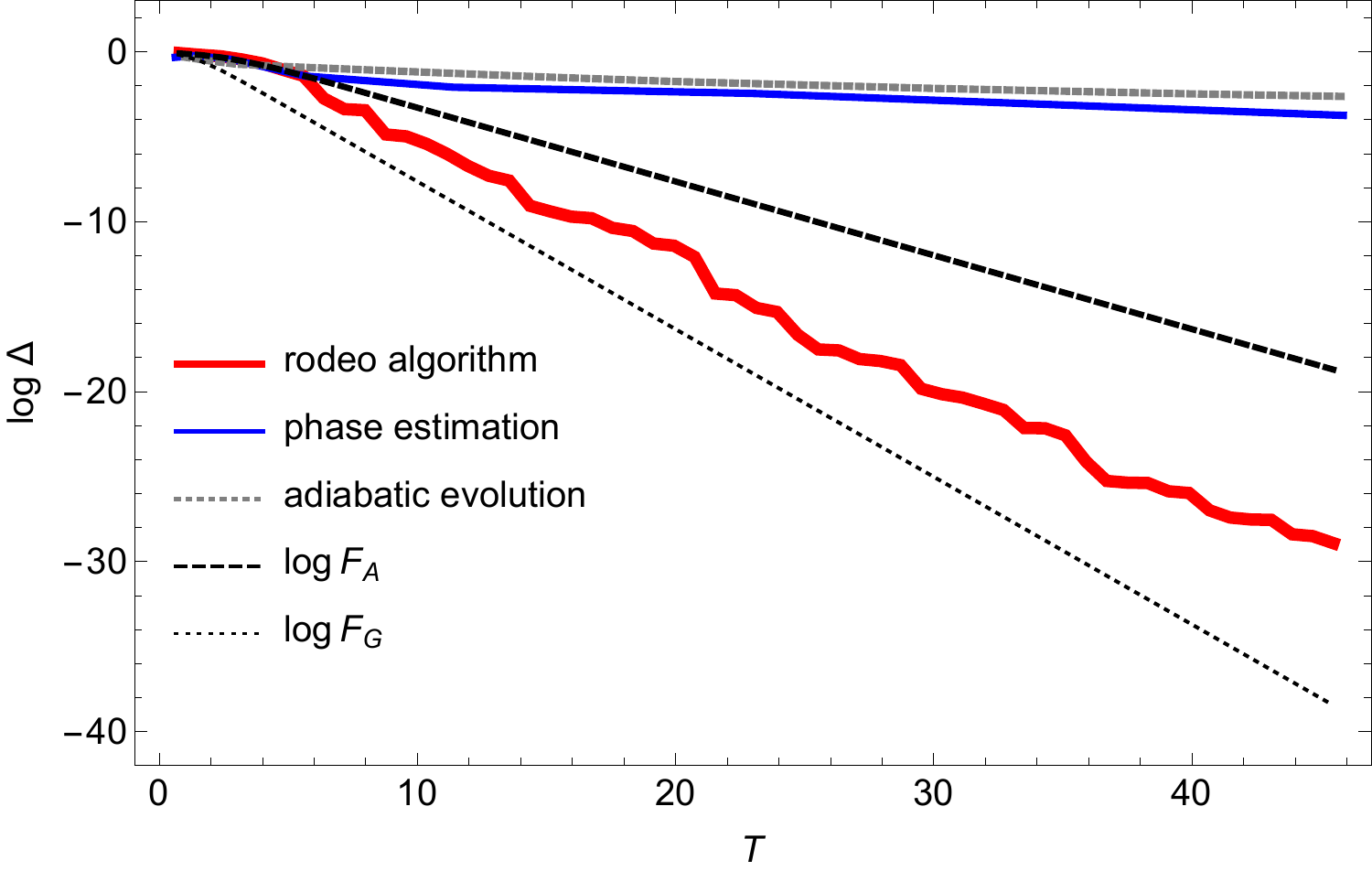}
\caption{(color online) {\bf Logarithm of the wave function error versus the total propagation time for the Heisenberg model.} We plot $\log \Delta$, versus the total propagation time, $T$, for the Heisenberg model. We show results for the rodeo algorithm, phase estimation, and adiabatic evolution.  We also show the asymptotic estimates $\log F_A$ and $\log F_G$.}
\label{Heisenberg_error_T}
\end{figure}

In Fig.~\ref{Heisenberg_error_T}, we show results for the logarithm of the wave function error for the rodeo algorithm when preparing the energy eigenstate $\ket{E_j}$ corresponding to $E_j = -18.1$.  We plot the logarithm of the magnitude of the orthogonal complement, $\log \Delta$, versus the total time evolution, $T$, for $t_{\rm RMS} = 1$.  For comparsion, we show the estimates $\log F_A$ and $\log F_G$ defined in Eq.~(\ref{estimates}).  As expected, for small $T$, $\log F_A$ provides a good estimate, while for very large $T$, $\log F_G$ provides a better estimate.  For comparison, we also show the analogous results obtained with the same initial state but instead using phase estimation and adiabatic evolution.  For the adiabatic evolution calculation, we use the initial Hamiltonian $H_I = \sum_{j=1}^{10} (-1)^j \sigma^z_j $,
with an interpolating function $H(t) = \cos^2[\pi t/(2T)]H_I + \sin^2[\pi t/(2T)]H_{\rm obj}$.  We note that phase estimation and adiabatic evolution are similar in performance, while the rodeo algorithm is exponentially faster than both.

Even though the rodeo algorithm is more efficient than adiabatic evolution for eigenstate preparation, we can use adiabatic evolution as a preconditioner for the rodeo algorithm, in order to amplify the overlap of the initial state with the desired eigenvector.  In Table~\ref{overlap}, we show the overlap probability for energy eigenvector $\ket{E_j}$ with $E_{j}=-18.1$ after preconditioning with adiabatic evolution for time $t_{\rm AE}$ and applying $N$ cycles of the rodeo algorithm with $t_{\rm RMS} = 5$.  We see that by preconditioning with $t_{\rm AE}=5$, we achieve a more than sevenfold increase in the initial state overlap probability.  We gain a significant computational advantage when using adiabiatic evolution as a preconditioner for the rodeo algorithm and expect this strategy to be very useful for larger system calculations.
\begin{table}[h]
\begin{center}
\caption{Overlap probability with energy eigenvector $\ket{E_j}$ with $E=E_{j}=-18.1$ after preconditioning with adiabatic evolution for time $t_{\rm AE}$ and the applying $N$ cycles of the rodeo algorithm using Gaussian random values for $t_n$ with $t_{\rm RMS} = 5$.\\}
\label{overlap}
\begin{tabular}{|c|c|c|c|c|c|}
\hline
$E_j$ & $t_{\rm AE}$ & $N=0$ & $N = 3$ & $N = 6$ & $N= 9$ \\
\hline
$ -18.1 $ & 0 & $  0.110 $ & $  0.746 $ & $  0.939 $ & $  0.997  $ \\
$ -18.1 $ & 5 & $  0.83074 $ & $ 0.99875 $ & $  0.99988 $ & $  0.99999  $ \\
\hline
\end{tabular}
\end{center}
\end{table}

As a second application of the rodeo algorithm, we consider the Anderson localization model in one dimension, which describes the transition between extended and localized electronic states in the presence of spatially-varying disorder \cite{Anderson:1958vr,Abrahams:1979zz}.  Our object Hamiltonian $H_{\rm obj}$ describes a single particle on a periodic, one-dimensional lattice with $100$ sites.  Let us denote the position basis states as $\ket{k}$ with $k=0,\cdots,99$.  The matrix elements of the object Hamiltonian are 
\begin{equation}
   [H_{\rm obj}]_{k',k}=-\delta_{k',k+1} - \delta_{k',k-1} + c_{k}\delta_{k',k},
\end{equation} 
where the coefficients $c_k$ provide diagonal disorder that control the amount of localization of the electronic wave functions.  In this example, we take the diagonal terms $c_k$ to be Gaussian random numbers with zero mean and root-mean-square value equal to $\tfrac{1}{2}$.  With this amount of diagonal disorder, the resulting orbitals are rather localized in space.  In the Supplemental Materials, we also present results for the case where diagonal disorder is much weaker and the energy eigenstates are delocalized in space.

\begin{figure}
\centering
\includegraphics[width=8.5cm]{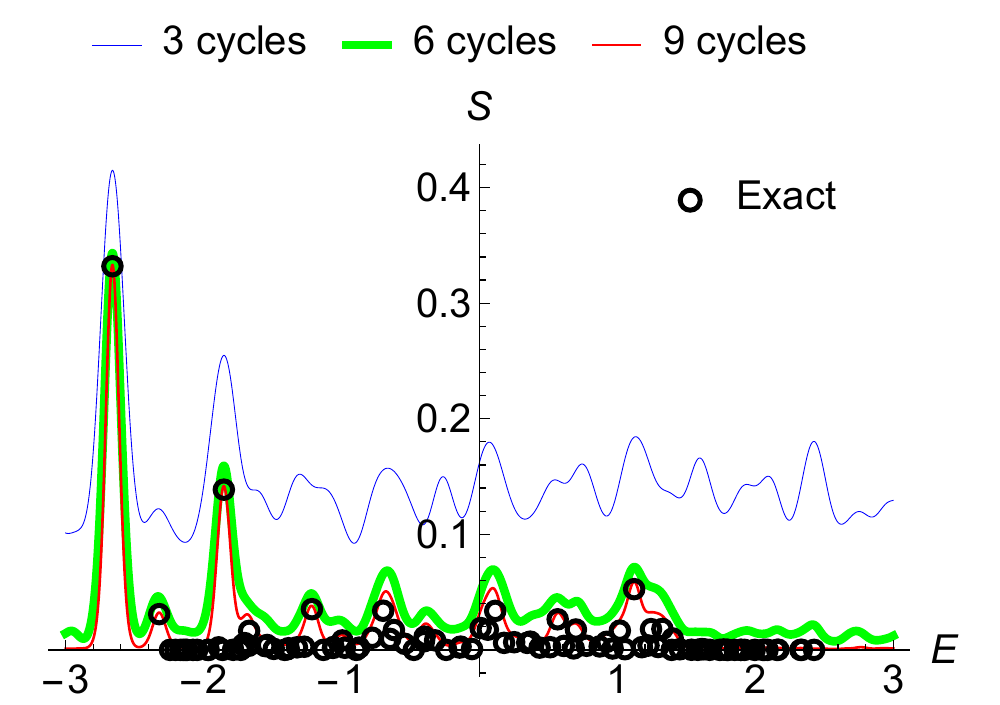}
\caption{(color online) {\bf Initial-state spectral function for the Anderson localization model with diagonal disorder $\tfrac{1}{2}$.} We plot the initial-state spectral function using the rodeo algorithm for the Anderson localization model with the root-mean-square diagonal disorder equal to $\tfrac{1}{2}$.  We show results for $3$ (thin blue line), $6$ (thick green line), and $9$ (medium red line) cycles. We have averaged over $20$ sets of Gaussian random values for $t_n$ with $t_{\rm RMS}=10$.  For comparison, we also show the exact initial-state spectral function with black open circles.}
\label{Anderson_spectrum}
\end{figure}

As a variational ansatz for the ground state, we take our initial state to be the basis state $\ket{k_{\rm min}}$ such that $c_{k_{\rm min}}$ is the lowest diagonal element of the Hamiltonian.  This choice reflects our physical intuition that the ground state is spatially localized.  In Fig.~\ref{Anderson_spectrum}, we plot the initial-state spectral function using the rodeo algorithm for the Anderson localization model with the root-mean-square diagonal disorder equal to $\tfrac{1}{2}$.  We consider $N=3$ (thin blue line), $6$ (thick green line), and $9$ (medium red line) cycles. We have also averaged over $20$ sets of Gaussian random values for $t_n$ with $t_{\rm RMS}=10$.  For comparison, we show the exact initial-state spectral function with black open circles.  We see that the exact spectral function is well reproduced.  The fact that our point-like initial state has significant overlap with only a small fraction of energy eigenstates is an indication of the localized character of the orbitals.

In this letter, we have presented a new method called the rodeo algorithm for preparing quantum eigenstates and determining spectral properties.  It uses a tunable energy filter and stochastic methods to prepare any eigenstate of a given quantum Hamiltonian and can in fact be used to prepare the eigenstates of any quantum observable.  It is an efficient algorithm with exponential convergence that can be performed using circuits with relatively short gate depth. In particular, the speed for eigenstate preparation is exponentially faster than that for phase estimation or adiabatic evolution. The rodeo algorithm can be combined with variational methods and/or quantum adiabatic evolution to provide a tool for solving the quantum many-body problem even when there is no {\it a priori} information about the target eigenvector.  It has the potential for wide applicability across many different fields, including combinatorial optimization problems, strongly-correlated electrons, nuclear structure and dynamics, and lattice quantum chromodynamics.  The results presented here were obtained from classical computation, but we are now working to implement the rodeo algorithm on quantum devices using several quantum Hamiltonians. The results will be presented in future publications.

\paragraph*{Acknowledgements}

We are grateful for discussions with Joseph Carlson, Gabriel Given, Erik Gustafson, Caleb Hicks, Ning Li, Bing-Nan Lu, Yannick Meurice, Sofia Quaglioni, John Rickert, Alessandro Roggero, Avik Sarkar, Nathan Wiebe, Kyle Wendt, and Boris Zbarsky.  We acknowledge financial support from the U.S. Department of Energy (DE-SC0018638 and DE-SC0021152), Los Alamos National Laboratory, NUCLEI SciDAC-4 collaboration, and the Center for Excellence in Education as part of the 2020 Research Science Institute.

\bibliography{References}

\pagebreak
\beginsupplement

\section{Supplemental Materials}

\subsection{Initial-state spectral function for the Heisenberg model}

\begin{figure}[h]
\centering
\includegraphics[width=8.5cm]{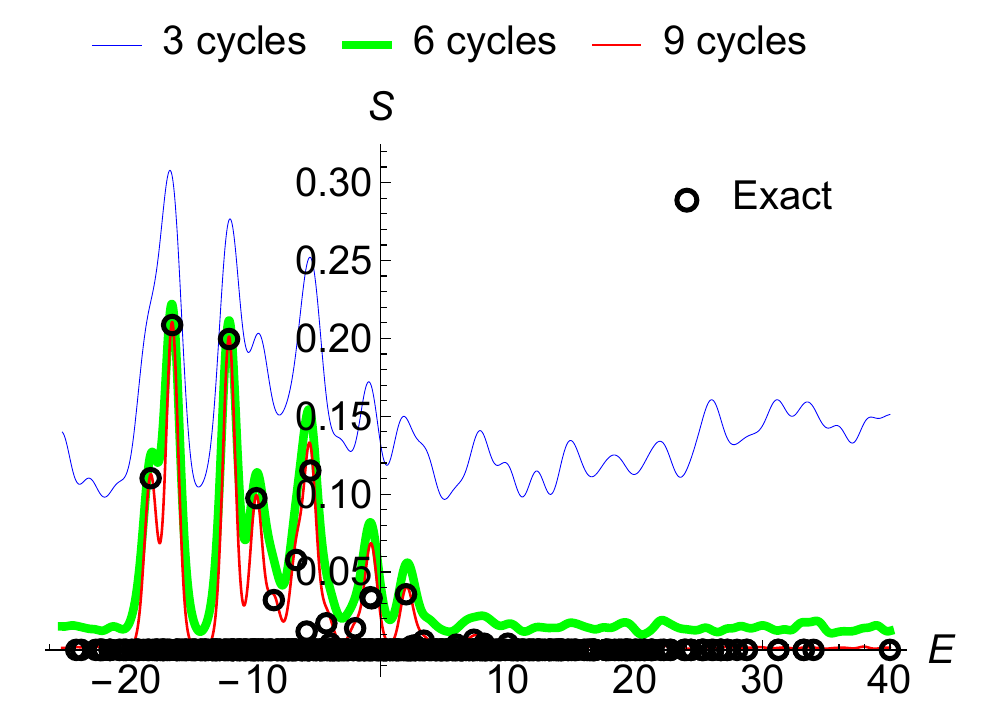}
\caption{{\bf Initial-state spectral function for the Heisenberg model using $t_{\rm RMS}=1$.} We plot the initial-state spectral function using the rodeo algorithm for the Heisenberg spin chain with $3$ (thin blue line), $6$ (thick green line), and $9$ (medium red line). We have averaged over $20$ sets of Gaussian random values for $t_n$ with $t_{\rm RMS}=1$.  For comparison, we also show the exact initial-state spectral function with black open circles.}
\label{Heisenberg_spectrum_1}
\end{figure}

In Fig.~\ref{Heisenberg_spectrum_1}, we plot the initial-state spectral function using the rodeo algorithm for the Heisenberg spin chain with $N=3$ (thin blue line), $6$ (thick green line), and $9$ (medium red line) cycles. We have averaged over $20$ sets of Gaussian random values for $t_n$ with $t_{\rm RMS}=1$.  For comparison, we also show the exact initial-state spectral function with black open circles.  We observe that the spectral function is well reproduced even though $t_{\rm RMS}=1$ is rather small and the number of cycles is also not large.

\subsection{Eigenstate preparation for the Heisenberg model}

In addition to computing the initial-state spectral function, we can also prepare any energy eigenstate that has nonzero overlap with our initial state. In Table~\ref{overlap} we show the overlap probability with energy eigenvector $\ket{E_j}$ after $N$ cycles of the rodeo algorithm.  We use Gaussian random values for $t_n$ with $t_{\rm RMS} = 5$ and set $E=E_{j}$.  All of the energy eigenvectors with nonzero overlap with our initial state can be prepared with a relatively small number of rodeo cycles. 

\begin{table}[h]
\begin{center}
\caption{Overlap probability with energy eigenvector $\ket{E_j}$ after $N$ cycles of the rodeo algorithm using Gaussian random values for $t_n$ with $t_{\rm RMS} = 5$ and $E=E_{j}$.}
\label{overlap}
\begin{tabular}{|c|c|c|c|c|}
\hline
$E_j$ & $N=0$ & $N = 3$ & $N = 6$ & $N= 9$ \\
\hline
$ -18.1 $ & $  0.110 $ & $  0.746 $ & $  0.939 $ & $  0.997  $ \\
$ -16.4 $ & $  0.209 $ & $  0.841 $ & $  0.993 $ & $  1.000  $ \\
$ -11.9 $ & $  0.200 $ & $  0.629 $ & $  0.889 $ & $  0.999  $ \\
$ -9.76 $ & $  0.0974 $ & $  0.488 $ & $  0.903 $ & $  0.999  $ \\
$ -8.38 $ & $  0.0320 $ & $  0.467 $ & $  0.832 $ & $  0.993  $ \\
$ -6.63 $ & $  0.0577 $ & $  0.309 $ & $  0.818 $ & $  0.996  $ \\
$ -5.81 $ & $  0.0118 $ & $  0.179 $ & $  0.637 $ & $  0.817  $ \\
$ -5.52 $ & $  0.115 $ & $  0.456 $ & $  0.766 $ & $  0.997  $ \\
$ -4.26 $ & $  0.0171 $ & $  0.144 $ & $  0.696 $ & $  0.995  $ \\
$ -3.95 $ & $  0.00401 $ & $  0.0430 $ & $  0.343 $ & $  0.952  $ \\
$ -2.00 $ & $  0.0139 $ & $  0.158 $ & $  0.593 $ & $  0.942  $ \\
$ -0.802 $ & $  0.0338 $ & $  0.216 $ & $  0.545 $ & $  0.594  $ \\
$ -0.704 $ & $  0.0331 $ & $  0.286 $ & $  0.540 $ & $  0.585  $ \\
$ 2.00 $ & $  0.0357 $ & $  0.371 $ & $  0.925 $ & $  0.994  $ \\
$ 2.42 $ & $  0.00235 $ & $  0.0122 $ & $  0.0874 $ & $  0.521  $ \\
$ 2.68 $ & $  0.00291 $ & $  0.0845 $ & $  0.639 $ & $  0.929  $ \\
$ 3.39 $ & $  0.00592 $ & $  0.0360 $ & $  0.754 $ & $  0.943  $ \\
$ 5.96 $ & $  0.00336 $ & $  0.0951 $ & $  0.559 $ & $  0.981  $ \\
$ 7.33 $ & $  0.00650 $ & $  0.184 $ & $  0.792 $ & $  0.978  $ \\
$ 8.13 $ & $  0.00393 $ & $  0.0832 $ & $  0.665 $ & $  0.841  $ \\
$ 8.24 $ & $  0.00105 $ & $  0.0275 $ & $  0.142 $ & $  0.289  $ \\
$ 10.0 $ & $  0.00397 $ & $  0.0128 $ & $  0.295 $ & $  0.902 $ \\

\hline
\end{tabular}

\end{center}
\end{table}

\subsection{Initial-state spectral function for the Anderson localization model}

\begin{figure}
\centering
\includegraphics[width=8.5cm]{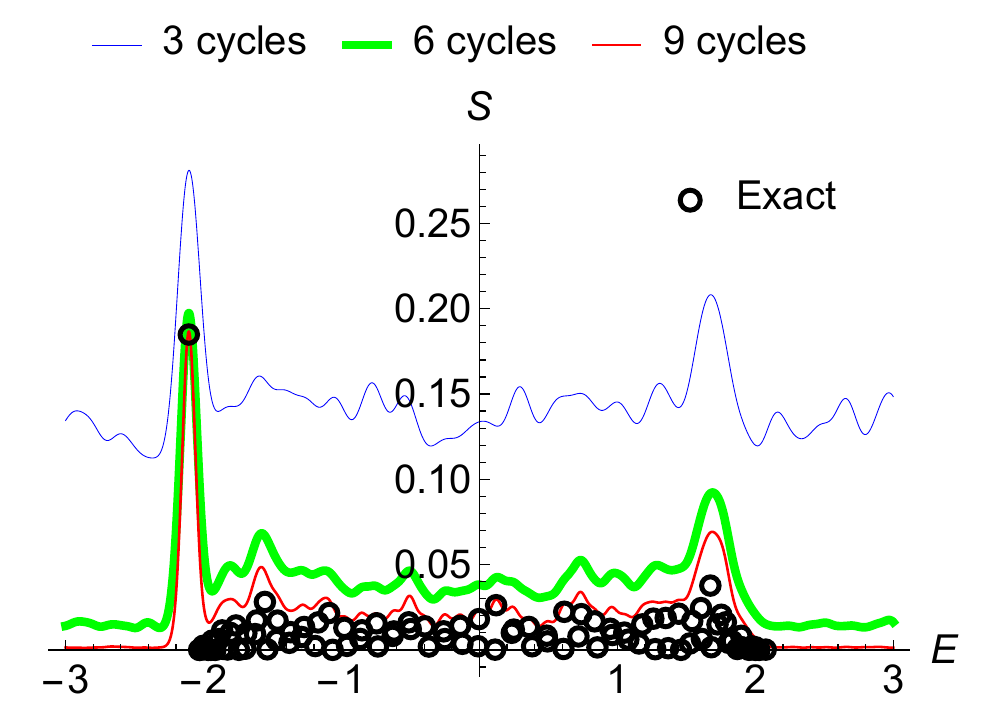}
\caption{(color online) {\bf Initial-state spectral function for the Anderson localization model with diagonal disorder $\tfrac{1}{8}$ and using $t_{\rm RMS}=10$.} We plot the initial-state spectral function using the rodeo algorithm for the Anderson localization model with the root-mean-square diagonal disorder equal to $\tfrac{1}{8}$.  We show results for $3$ (thin blue line), $6$ (thick green line), and $9$ (medium red line) cycles. We have averaged over $20$ sets of Gaussian random values for $t_n$ with $t_{\rm RMS}=10$.  For comparison, we also show the exact initial-state spectral function with black open circles.}
\label{Anderson_weak_spectrum_10}
\end{figure}

As a final example, we consider the Anderson localization model again, but this time with much weaker diagonal disorder.  In Fig.~\ref{Anderson_weak_spectrum_10}, we plot the initial-state spectral function using the rodeo algorithm for the Anderson localization model with the root-mean-square diagonal disorder equal to $\tfrac{1}{8}$.  With this small amount of diagonal disorder, the excited state orbitals are delocalized in space.  We again take our initial state to be the basis state $\ket{k_{\rm min}}$ such that $c_{k_{\rm min}}$ is the lowest diagonal element of the Hamiltonian. We show $N=3$ (thin blue line), $6$ (thick green line), and $9$ (medium red line) cycles. We have averaged over $20$ sets of Gaussian random values for $t_n$ with $t_{\rm RMS}=10$.  For comparison, we also show the exact initial-state spectral function with black open circles.  We observe that the spectral function is again well reproduced.  The fact that our point-like initial state has a small but discernible overlap with most of the excited states is evidence of the extended spatial character of the orbitals.

\begin{figure}
\centering
\includegraphics[width=8.5cm]{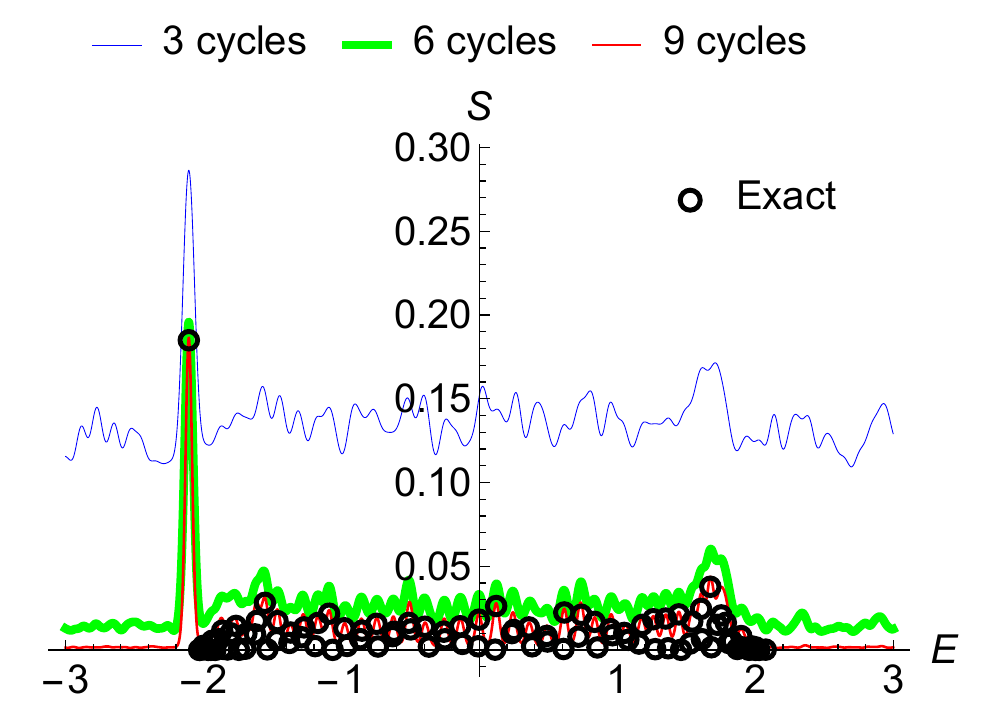}
\caption{{\bf Initial-state spectral function for the Anderson localization model with diagonal disorder $\tfrac{1}{8}$ and using $t_{\rm RMS}=20$.} We plot the initial-state spectral function using the rodeo algorithm for the Anderson localization model with the root-mean-square diagonal disorder equal to $\tfrac{1}{8}$.  We show results for $3$ (thin blue line), $6$ (thick green line), and $9$ (medium red line) cycles. We have averaged over $20$ sets of Gaussian random values for $t_n$ with $t_{\rm RMS}=20$.  For comparison, we also show the exact initial-state spectral function with black open circles.}
\label{Anderson_weak_spectrum_20}
\end{figure}

While keeping the root-mean-square diagonal disorder equal to $\tfrac{1}{8}$, we now show the corresponding results using $t_{\rm RMS}=20$ to better resolve the details of the excited-state spectrum.
In Fig.~\ref{Anderson_weak_spectrum_20}, we plot the initial-state spectral function using the rodeo algorithm for the Anderson localization model, again with the initial state chosen to be the basis state $\ket{k_{\rm min}}$ such that $c_{k_{\rm min}}$ is the lowest diagonal element of the Hamiltonian. We show $N=3$ (thin blue line), $6$ (thick green line), and $9$ (medium red line) cycles. We have averaged over $20$ sets of Gaussian random values for $t_n$ with $t_{\rm RMS}=20$.  We take this larger value for $t_{\rm RMS}=20$ to resolve the details of the excited-state spectrum.  For comparison, we also show the exact initial-state spectral function with black open circles.  We see that the entire spectral function is well reproduced.  Our point-like initial state has a small but discernible overlap with most of the excited states, indicating the extended spatial character of the excited-state orbitals.  

\end{document}